\documentclass{article}
\usepackage{amsmath, amssymb, graphics, setspace}

\newcommand{\mathsym}[1]{{}}
\newcommand{\unicode}[1]{{}}

\begin{document}

\title{Strain mediated tri- and quattro- interactions of adatoms.}
\author{Wolfgang Kappus\\
\\
wolfgang.kappus@t-online.de}
\date{v0: 2016-10-23}
\maketitle

\section*{Abstract}

Lateral interactions of oxygen adatoms derived from first-principles calculations of the O-Pd(100) system had been claimed to include trio- and quattro
terms beside pair interactions. This paper is dedicated to extend an earlier model for substrate strain mediated interactions between adatom pairs
to include trio- and quattro terms. While monomers (sitting on high symmetric sites) are supposed to create isotropic stress on the substrate, dimers
would create anisotropic stress. The requirement of mechanical equilibrium allows to formulate the elastic energy as a sum of pair-, trio- and quattro
terms, solved by respective an eigenvalue problems. Resulting interactions are strongly angle dependent and also reflect the elastic anisotropy of
the substrate. A continuum elastic model supports the long range character of the elastic interaction but should also describe some short range features.
For the elastically isotropic tungsten substrate a closed solution for quattro interactions is presented, for trio inter cations numerical values
are given.

\section*{1.Introduction}

Interactions of adatoms are a subject of continuous interest, various different interaction mechanisms have been described in detail [1]. Strain
mediated adatom interactions with the basic \textup{ \(s^{-3}\)} distance proportionality and the strong influence of the substrates elastic anisotropy
were discussed already in the 1970 decade based on elastic continuum theory [2,3,4] . Also a cutoff method was proposed [3] to handle the \textup{
\(s^{-3}\)} singularity, leading to an attractive interaction at very short range. { } { }

Elastic interactions appear relevant since many adatom structures reflect the elastic anisotropy of the substrate. Direct measurements of elastic
interactions are difficult, so theoretical models seem attractive. Lateral interactions of oxygen adatoms derived from first-principles calculations
of the O-Pd(100) system were claimed to include trio- and quattro terms beside pair interactions [5] which offers an attractive challenge for testing
the reach of elastic models. In a parallel study the first-principles calculations are interpreted using long-range strain mediated interactions
[6]. The current analysis is dedicated to extend the earlier model for substrate strain mediated interactions between adatoms [3] to include trio-
and quattro terms. 

In the pair model the interaction of adatoms is mediated by their strain fields generated by single adatoms exerting isotropic stress to their vicinity.
Isotropic stress is a consequence of adatom locations on sites with 3- or 4-fold symmetry, selected for simplicity. In a lattice description adatoms
would exert forces to their immediate substrate neighbors creating a displacement field equivalent to a strain field. { }

Basic idea of the current model are adatom pairs stretching or compressing their bond due to their position on substrate sites. Those pairs exert
forces to their immediate substrate neighbors and thus create a displacement field to balance the forces. The strain fields of such pairs will mediate
interactions between pairs and also between pairs and monomers. The interaction of a monomer with a pair is a trio interaction, the interaction of
two pairs is a quattro interaction. Lattice mismatch - known as source of substrate strain [7] - is the mesoscopic consequence of such pair interactions.

If forces between pairs are central, the model can be kept simple and the resulting stress fields can (with symmetry restrictions) be described with
one more free parameter than the previous monomer interaction model. The requirement of mechanical equilibrium allows to formulate the elastic energy
as a sum of pair-, trio- and quattro terms and to solve them separately. { } { } 

Calculations below are based on a plain wave expansion of the displacement field and an eigenvalue method, introduced in [8], equivalent to a Green
function method used in other papers [7]. Closely following [3] the elastic energy is diagonalized in a set of elastic modes and a generalized eigenvalue
problem is established and solved. The elastic interaction will show up as a series of special functions.

The paper is organized as follows:\\
After outlining the motivation in section 1 the elastic model and its proof are given in section 2. In section 3 the method for calculating interactions
is detailed. In section 4 applicability and limitations of the model are discussed. Section 5 closes with a summary of the results.

\section*{2.Model details}

In this section the overall energy is introduced and the elastic part is diagonalized in terms of eigenmodes. A set of linear equations for the equilibrium
displacements in the bulk and at the surface is deduced with an eigenvalue $\omega $ to be used later for calculating the elastic interaction of
adatoms and dimers. The tensor (and tensor product) notation used below (bold letters) tries to concentrate on the essentials, details of the indices
are given in [3]. The text in the following follows close to [3] but needs extensions to handle the difference between adatom distributions, responsible
for pair interactions, and the distributions of adatom pairs, responsible for trio- and quattro interactions. The subscripts\textit{  }\textit{ k,l}
will be used to label different distributions. The subscript $\lambda $ will be used to label eigenmodes.

\subsection*{2.1.Surface energy }

The adatom surface energy H is assumed to consist of an elastic part \(H_{\text{el}}\) and part \(H_{\text{other}}\) comprising other interactions
{ } { } { } { }

\[H=H_{\text{el}}+H_{\text{other}} .\text{                                                         }(2.1)\]

In the following sections the focus will be put on elastic interactions comprising also short range.

\subsection*{2.2.Elastic energy }

Adatoms exert forces on their substrate neighbors leading to a displacement of substrate atoms to balance those forces. Such displacements will increase
or decrease the energy of the system. In a continuum description adatoms exert stress parallel to the surface leading to substrate strain which in
turn can lead to an attraction or a repulsion of neighboring adatoms. The strength of such strain mediated interaction will depend on the stress
adatoms exert on the substrate and on the stiffness of the substrate. 

The elastic energy of a substrate with adatoms in a continuous description is given by the sum of two parts, the energy of the distorted substrate
and the energy of adatoms exerting tangential forces on the substrate { }

\[H_{\text{el}} = \frac{1}{2} \int _V\pmb{\text{\textit{$\epsilon (r)$}}\text{\textit{$ $}}c\text{\textit{$ $}}\text{\textit{$\epsilon (r)$}}\text{\textit{$dr$}}}+\int
_S\pmb{\text{\textit{$\epsilon (s)$}}}\pmb{\text{\textit{$\pi (s)$}}}\pmb{\text{\textit{$ds$}}} .\text{                                      }(2.2)\]

Here \pmb{ \textit{ $\epsilon $}}=[\(\epsilon _{\alpha \beta }\)] denotes the strain tensor, \pmb{ \textit{ c}}=[\(c_{\alpha \beta \mu \nu }\)] denotes
the elastic constants tensor, and \pmb{ \textit{ $\pi $}}=[\(\pi _{\mu \nu }\)] denotes the stress tensor. The integrals comprise the bulk V or the
surface S. The strain field \pmb{ \textit{ $\epsilon $(r)}} is related to the displacement field \pmb{ \textit{ u(r)}} by

\[\epsilon _{\alpha \beta }(\pmb{r})=\frac{1}{2} \left(\nabla _{\alpha }u_{\beta }(\pmb{r})+\nabla _{\beta } u_{\alpha }(\pmb{r})\right) .\text{
                                        }(2.3)\]

If adatoms do not interact directly with their neighbors, the stress field \pmb{ \textit{ $\pi $(s)}} depends only on the configuration of single
adatoms. In case where adatoms do interact with their next neighbors directly, the stress field \pmb{ \textit{ $\pi $(s)}} depends also on the configuration
of adatom pairs, because pairs exert forces to the substrate different from monomers. Introducing \textit{ k} as indicator of \textit{ np} different
cases, an adatom configuration creates a stress field \pmb{ \textit{ $\pi $(s)}} superimposed of different types \pmb{ \textit{ \(\pi _k\)}}\pmb{
\textit{ (s)}}

\[\pmb{\pi \text{\textit{$($}}s\text{\textit{$)$}}}= \sum _{k=1}^{\text{np}}  \pmb{\pi }_k\pmb{\text{\textit{$($}}}\pmb{s}\pmb{\text{\textit{$)$}}}=\sum
_{k=1}^{\text{np}}  \pmb{\text{\textit{$P_k$}}}\rho _k(\pmb{s})\text{  },\text{                                          }(2.4)\]

where we have introduced \textit{ np} different types of { }force dipole / stress tensors \(\pmb{\text{\textit{$P_k$}}}\) and { }densities \(\rho
_k(\pmb{s})\) of adatom monomers or pairs. On (100) surfaces with adatom positions of fourfold coordinated sites \textit{ np}=3; \(\pmb{\text{\textit{$P_1$}}}\)=\(P_1\)[\(\delta
_{\alpha \beta }\)] stand for an isotropic monomer stress tensor, \(\pmb{\text{\textit{$P_2$}}}\)=\(P_2\)[\(\delta _{\text{$\alpha $1}}\)\(\delta
_{\text{$\beta $1}}\)] and \(\pmb{\text{\textit{$P_3$}}}\)=\(P_3\)[\(\delta _{\text{$\alpha $2}}\)\(\delta _{\text{$\beta $2}}\)] for anisotropic
dimer stress tensors. \(\rho _1(\pmb{s})\) then stands for the adatom monomer density distribution, \(\rho _2(\pmb{s})\) for the density distribution
of x-directed adatom pairs and \(\rho _3(\pmb{s})\) for the density distribution of y-directed adatom pairs. \\
With Eq. (2.4) Eq. (2.2) reads

\[H_{\text{el}} = \frac{1}{2}\text{  }\int _V \pmb{\text{\textit{$\epsilon (r)$}}\text{\textit{$ $}}c\text{\textit{$ $}}\text{\textit{$\epsilon (r)$}}\text{\textit{$dr$}}}\underline{
}+ \sum _{k=1}^{\text{np}} \int _S \pmb{\text{\textit{$\epsilon (s)$}}} \pmb{\text{\textit{$P_k$}}}\rho _k(\pmb{s}) \pmb{\text{\textit{$ds$}}}\text{
 }.\text{                                                }(2.5)\]

The strain field \pmb{ \textit{ $\epsilon $(r)}} is determined for given densities \(\rho _k(\pmb{s})\) by the requirement of mechanical equilibrium

\[\delta  \left.H_{\text{el}}\right/\delta  u_{\alpha }\pmb{\text{\textit{$($}}}\pmb{\text{\textit{$ $}}}\pmb{r})\underline{ } = 0 ,\text{      
                                                              }(2.6)\]

which leads to freedom of stress in the bulk

\[\text{\textit{$\pmb{\nabla \nabla c} \pmb{u(r)}$}}= \pmb{0}\pmb{\text{\textit{$ $}}},\text{      }\pmb{r}\text{\textit{   }}\text{in}\text{  }V\text{
                                        }(2.7)\text{    }\]

and freedom of stress normal \pmb{ \textit{ n(s)}} to the surface 

\[\pmb{\text{\textit{$ $}}}\pmb{n\text{\textit{$($}}\text{\textit{$s$}}\text{\textit{$)$}}\text{\textit{$\nabla c$}}\text{\textit{$ $}}\text{\textit{$u(s)$}}}=
\sum _{k=1}^{\text{np}}  \pmb{\text{\textit{$P_k$}}}\pmb{\text{\textit{$\nabla $}}} \rho _k(\pmb{s}) .\text{    }\pmb{s}\text{\textit{   }}\text{on}\text{
 }S.\text{        }(2.8)\text{   }\]

With the help of Eqs. (2.7) and (2.8) Eq. (2.5) can be rewritten neglecting boundary effects 

\[H_{\text{el}}=-\frac{1}{2} \sum _{k=1}^{\text{np}}  \int _S \pmb{\text{\textit{$u(s)$}}} \pmb{\text{\textit{$P_k$}}}\pmb{\text{\textit{$\nabla
$}}} \rho _k(\pmb{s}) \pmb{\text{\textit{$ds$}}} .\text{                }(2.9)\]

\subsection*{2.3.Elastic interaction }

An elastic interaction between adatoms can be derived from Eq. (2.9) with an ansatz relating adatom density and displacement field { }at the surface

\[\pmb{u}(\pmb{s}\pmb{)}=\sum _{l=1}^{\text{np}} \left.\pmb{\pmb{u}_l}\right(\pmb{s}\pmb{)}= -\sum _{l=1}^{\text{np}}  \int _S \pmb{\text{\textit{$A\left(s,s'\right)$}}}
\pmb{\text{\textit{$P_l$}}}\pmb{\pmb{\nabla }'}\rho _l\left(\pmb{\pmb{s}'}\right) \pmb{\text{\textit{$ds'$}}} ,\text{               }(2.10)\]

which gives inserted in Eq. (2.9)

\[H_{\text{el}}=\frac{1}{2} \sum _{k=1}^{\text{np}} \sum _{l=1}^{\text{np}}  \int _S \int _S \pmb{\text{\textit{$P_k$}}}\pmb{\nabla }\rho _k(\pmb{s})
\pmb{A}\left(\pmb{s},\pmb{\text{\textit{$s'$}}}\right) \pmb{\text{\textit{$P_l$}}}\pmb{\nabla '}\rho _l\left(\pmb{\pmb{s}'}\right) \pmb{\text{\textit{$ds$}}}
\pmb{\text{\textit{$ds'$}}}.\text{                                 }(2.11)\text{                         }\]

With

\[V_{kl}\left(\pmb{s},\pmb{\text{\textit{$s'$}}}\right)= \pmb{\text{\textit{$P_k$}}} \pmb{\text{\textit{$P_l$}}}\pmb{\nabla } \pmb{\nabla '}\pmb{A}\left(\pmb{s},\pmb{\text{\textit{$s'$}}}\right)\text{
                                                                       }(2.12)\text{                         }\]

this can be written

\[H_{\text{el}}=\frac{1}{2} \sum _{k=1}^{\text{np}} \sum _{l=1}^{\text{np}}  \int _S \int _S \rho _k(\pmb{s}) V_{\text{\textit{kl}}}\left(\pmb{s},\pmb{\text{\textit{$s'$}}}\right)
\rho _l\left(\pmb{\pmb{s}'}\right) \pmb{\text{\textit{$ds$}}} \pmb{\text{\textit{$ds'$}}}\pmb{\text{\textit{$,$}}}\text{                        
       }(2.13)\text{                         }\]

with \(V_{\text{\textit{kl}}}\left(\pmb{s},\pmb{\text{\textit{$s'$}}}\right)\) denoting the elastic interaction between adatoms (or pairs) of type
\textit{ k} at \pmb{ \textit{ s}} and adatoms (or pairs) of type \textit{ l} at \(\pmb{\text{\textit{$s'$}}}\). 

Fig.1 shows different combinations of adatoms or dimers and interactions. 1.a. monomers with stress tensor\(\pmb{ }\pmb{P}_1\) and interaction via
\(V_{11}\). 1.b. pairs with stress tensor \(\pmb{P}_2\) { }interacting via \(V_{22}\). 1.c. pairs with stress tensors \(\pmb{P}_2\) and \(\pmb{P}_3\)
{ }interacting via \(V_{23}\). 1.d. monomer with stress tensor \(\pmb{P}_1\) { }and pair with stress tensor \(\pmb{P}_2\) interacting via \(V_{12}\).

\subsection*{2.4.Eigenfunctions }

The interactions \(V_{\text{\textit{kl}}}\left(\pmb{s},\pmb{\text{\textit{$s'$}}}\right)\) are diagonalized by eigenfunctions \(X_{\lambda }\)(\pmb{
s}) and eigenvalues \(\omega _{\text{$\lambda $kl}}\) given by

\[\int _S V_{\text{\textit{kl}}}\pmb{\text{\textit{$($}}}\pmb{\text{\textit{$s,s'$}}}\pmb{\text{\textit{$)$}}} X_{\lambda }\left(\pmb{\pmb{s}'}\right)
\pmb{\text{\textit{$ds'$}}}=\text{  }\omega _{\text{\textit{$\lambda $kl}}}\text{  }X_{\lambda }(\pmb{s}) .\text{               }(2.14)\]

The eigenfunctions form a complete orthonormal basis

\[\text{  }\sum _{\lambda } X_{\lambda } (\pmb{s})\text{  }X_{\lambda }\left(\pmb{\pmb{s}'}\right) =\delta \left(\pmb{s}-\pmb{s}^{\pmb{'}}\right)
,\text{  }\int _S\text{  }X_{\lambda } (\pmb{s})\text{  }X_{\text{\textit{$\lambda '$}}}(\pmb{s})\text{  }\pmb{\text{\textit{$ds$}}} =\text{  }\delta
_{\text{\textit{$\lambda \lambda '$}}} .\text{                                }(2.15)\]

Expanding

\[\left.\rho _k\right(\pmb{s}\pmb{)}=\sum _{\lambda }  \rho _{\text{\textit{$\lambda $k}}} X_{\lambda } (\pmb{s})\text{                         
                           }(2.16)\]

we find from Eqs. (2.13), (2.16), (2.14), (2.15)

\[H_{\text{el}}=\frac{1}{2} \sum _{k=1}^{\text{np}} \sum _{l=1}^{\text{np}} \sum _{\lambda } \omega _{\text{\textit{$\lambda $k}}l}\text{  }\rho
_{\text{\textit{$\lambda $k}}} \rho _{\text{\textit{$\lambda $l}}} .\text{                                       }(2.17)\text{                  
      }\]

The eigenfunctions \(X_{\lambda }\) are called density modes. Each mode $\lambda $ and each density \(\left.\rho _k\right(\pmb{s}\pmb{)}\) causes
its own displacement field

\[\left.\pmb{u}_k\right(\pmb{s}\pmb{)}=\sum _{\lambda }  \rho _{\text{\textit{$\lambda $k}}} \pmb{u}_{\text{\textit{$\lambda $k}}} (\pmb{s}).\text{
                                                    }(2.18)\]

From Eqs. (2.14), (2.12), (2.10) { }we find

\[\pmb{\text{\textit{$P_k$}}}\pmb{\nabla }\pmb{u}_{\text{\textit{$\lambda $l}}} (\pmb{s})=\text{  }\omega _{\text{\textit{$\lambda $kl}}} X_{\lambda
}(\pmb{s}) .\text{                                                  }(2.19)\]

Which gives from Eq. (2.8)

\[\pmb{\text{\textit{$ $}}}\pmb{n\text{\textit{$($}}\text{\textit{$s$}}\text{\textit{$)$}}\text{\textit{$\nabla c\pmb{u}_{\text{$\lambda $l}}(s)$}}}\pmb{\text{\textit{$-$}}}\omega
_{\text{\textit{$\lambda $k}}l}^{-1} \pmb{\text{\textit{$P_k$}}}\pmb{\text{\textit{$P_l$}}}\pmb{\text{\textit{$\nabla $}}} \pmb{\text{\textit{$\nabla
$}}} \pmb{u}_{\text{\textit{$\lambda $l}}}(\pmb{s}) =\pmb{0}\pmb{ }.\text{              }\pmb{s}\text{\textit{   }}\text{on}\text{  }S.\text{   
    }(2.20)\text{   }\]

Eq. (2.20) together with the bulk condition (2.7), for a single mode written

\[\text{\textit{$\pmb{\nabla }\pmb{\nabla }\pmb{cu}_{\text{$\lambda $l}}\pmb{(r)}$}}=\pmb{0}\pmb{\text{\textit{$ $}}},\text{                    
                     }\pmb{r}\text{\textit{   }}\text{in}\text{  }V\text{                                         }(2.21)\text{    }\]

constitutes the generalized eigenvalue problem for the displacement field \(\pmb{u}_{\text{$\lambda $k}}\)(\pmb{ s}) associated with the density
mode { }\(X_{\lambda }\)(\pmb{ s}) according to Eq. (2.19). { }It turns out from Eqs. (2.20) and (2.21) that \(\pmb{u}(\pmb{s}\pmb{)}\) as given
in Eq. (2.10) with

\[A\left(\pmb{s},\pmb{\text{\textit{$s'$}}}\right)= \sum _{k=1}^{\text{np}} \sum _{l=1}^{\text{np}} \sum _{\lambda } \omega _{\text{\textit{$\lambda
$kl}}}^{-1} \pmb{u}_{\text{\textit{$\lambda $l}}}(\pmb{s}) \pmb{u}_{\text{\textit{$\lambda $k}}} \left(\pmb{s'} \right)\text{                   
                        }(2.22)\text{                         }\]

is a solution of Eqs. (2.7) and (2.8). The interaction Eq. (2.12) is then with Eq. (2.19)

\[V_{kl}\left(\pmb{s},\pmb{\text{\textit{$s'$}}}\right)=\sum _{\lambda } \omega _{\text{\textit{$\lambda $kl}}}\text{  }X_{\lambda }(\pmb{s}) X_{\lambda
} \left(\pmb{s'} \right) .\text{                                        }(2.23)\text{                         }\]

The interaction \(V_{kl}\left(\pmb{s},\pmb{\text{\textit{$s'$}}}\right)\) in Eq. (2.23) will in the following section be evaluated by associating
the density modes with plane waves. { }

\section*{3.Calculation of eigenvalues and interactions}

For the calculation of eigenvalues and interactions the label $\lambda $ will now be replaced by the wave vector \pmb{ $\kappa $ }of plane waves
for all combinations of tensors \(\pmb{P}_k\) and \(\pmb{P}_l\). \(\omega _{\text{$\lambda $kl}}\) now reads \(\omega _{\text{\textit{kl}}}(\pmb{\kappa
})\) and the density modes read

\[\pmb{X_{\text{\textit{$\lambda $}}}(s)}=\pmb{X_{\kappa }(s)}=S^{-1/2} \exp (\pmb{\text{\textit{$ $}}}i\pmb{\text{$\kappa $s}}) .\text{        
                                                  }(3.1)\]

\subsection*{3.1.Eigenvalue equation }

To solve Eqs. (2.20) and (2.21) the displacement fields \pmb{ \textit{ u(r)}} are expanded in plain waves separately for all combinations of tensors
{ }\(\pmb{P}_k\) and \(\pmb{P}_l\) 

\[\pmb{u}_{\text{\textit{kl}}}\pmb{\text{\textit{$($}}}\pmb{\kappa \text{\textit{$,$}}r}\pmb{\text{\textit{$)$}}}=\sum _{m=1}^3  A_{m,\text{\textit{kl}}}
(\pmb{\text{\textit{$\kappa $}}})\pmb{\text{\textit{$a_m$}}}(\pmb{\text{\textit{$\kappa $}}})\exp \left(\pmb{\text{\textit{$g_m$}}\text{\textit{$
$}}r}\right) ,\text{                                                            }(3.2)\]

where { }

\[\left.\pmb{g}_m\pmb{\text{\textit{$($}}}\pmb{\kappa }\right)= i\pmb{\kappa }+\pmb{n}q_m\text{                                                 
                                  }(3.3)\]

denotes the sum of wave vector \pmb{ \textit{ $\kappa $ }} parallel to the surface and a normal component. The \(\pmb{\text{\textit{$a_m$}}}\) are
normalized eigenvectors. The label \textit{ m} denotes 3 displacement modes. For a given \pmb{ \textit{ $\kappa $}} Eq. (2.21) is a homogeneous linear
equation for \(\pmb{\text{\textit{$a_m$}}}\)

\[\pmb{\left[c\text{\textit{$ $}}\text{\textit{$g_m$}}\text{\textit{$ $}}\text{\textit{$g_m$}}\text{\textit{$ $}}\text{\textit{$]$}}\right.\text{\textit{$a_m(\kappa
)$}}}=\pmb{0}\pmb{,}\text{                                                                       }(3.4)\]

where \(q_m\) are roots of the characteristic polynomial in \textit{ q}. The boundary conditions Eq. (2.20) become

\[\left[\pmb{c\text{\textit{$ $}}n\text{\textit{$ $}}\text{\textit{$g_m$}}\text{\textit{$ $}}\text{\textit{$a_m(\kappa )$}}}-\omega _{\text{\textit{kl}}}^{-1}(\pmb{\kappa
}) \pmb{\text{\textit{$P_k$}}} \pmb{\text{\textit{$P_l$}}} \pmb{\text{\textit{$g_m$}}}\pmb{\text{\textit{$ $}}}\pmb{\text{\textit{$g_m$}}}\pmb{\text{\textit{$
$}}}\pmb{\text{\textit{$a_m(\kappa )$}}}\right]A_{m,\text{\textit{kl}}}(\pmb{\kappa })=\pmb{0}\pmb{\text{\textit{$ $}}} ,\text{                 
     }(3.5)\text{        }\]

systems of homogeneous linear equations for \(A_{m,\text{\textit{kl}}}\)(\pmb{ \textit{ $\kappa $}}) with \(\omega _{\text{\textit{kl}}}^{-1}\)(\pmb{
\textit{ $\kappa $}}) as parameter, \(\pmb{\text{\textit{$g_m$}}}\) taken from the solution of Eq. (3.4), and \textit{ k},\textit{ l} standing for
the types of strain fields created by the stress { }tensors \(\pmb{\text{\textit{$P_k$}}}\), \(\pmb{\text{\textit{$P_l$}}}\) . Those eigenvalues
\(\omega _{\text{\textit{kl}}}^{-1}\)(\pmb{ \textit{ $\kappa $}}) finally define the interaction \(V_{\text{\textit{kl}}}\left(\pmb{s},\pmb{\text{\textit{$s'$}}}\right)\)
given in Eq. (2.23)

\[V_{\text{\textit{kl}}}\left(\pmb{s},\pmb{\text{\textit{$s'$}}}\right)=(2 \pi )^{-2}\int \omega _{\text{\textit{kl}}}(\pmb{\kappa }) \exp \left(i\pmb{\kappa
}\left(\pmb{s}-\pmb{\text{\textit{$s'$}}}\right)\right)d\pmb{\kappa } .\text{                                       }(3.6)\]

\subsection*{3.2.Cosine expansion and integration}

To make further use of the results of Eq. (3.6) \(\omega _{\text{\textit{kl}}}(\pmb{\kappa })\) is expanded in terms of a cosine series 

\[\omega _{\text{\textit{kl}}}(\pmb{\kappa })=\left| \pmb{\kappa }\right| \text{   }\sum _{p=0}^n \omega _{\text{\textit{kl}},p} \cos (p \phi )\text{
                                                               }(3.7)\]

where $\phi $ denotes the direction angle between the x- and the \pmb{ \textit{ $\kappa $}}-direction on (100) surfaces. For symmetry reasons \textit{
p} assumes on (001) surfaces the values 0,4,8 in the case of monomer adatoms sitting on fourfold coordinated substrate sites and the values of 0,2,4
in the case of pairs located on symmetry planes. Higher \textit{ p} values are truncated. The interaction Eq. (3.6) reads with Eq. (3.7)

\[V_{\text{\textit{kl}}}\left(\pmb{s},\pmb{\text{\textit{$s'$}}}\right)=(2 \pi )^{-2} \sum _{p=0}^n \omega _{\text{\textit{kl}},p} \int \kappa ^2d\kappa
 \int \cos (p \phi ) \exp (\text{i$\kappa $s} \cos (\chi -\phi ))d\phi \text{  },\text{                       }(3.8)\]

where \textit{ s}=Abs(\pmb{ \textit{ s}}-\(\pmb{\text{\textit{$\pmb{s}'$}}}\)) and $\chi $ is the angle between the x- and the \pmb{ \textit{ s}}-\(\pmb{\text{\textit{$\pmb{s}'$}}}\)direction.
Performing the $\phi $ integral we get

\[V_{\text{\textit{kl}}}\left(\pmb{s},\pmb{\text{\textit{$s'$}}}\right)=(2 \pi )^{-1} \sum _{p=0}^n \omega _{\text{\textit{kl}},p} \cos (p \chi )
\cos (p \pi /2)\int \kappa ^2J_p(\kappa  s)d\kappa \text{  },\text{                                     }(3.9)\]

where \(J_p\) are Bessel functions of the order \textit{ p}. The $\kappa $ integral diverges. We restrict the \textit{ $\kappa $}\pmb{ \textit{ 
}} values to the first Brillouin zone by introducing a smooth cutoff function of the order of the inverse lattice constant 

\[V_{\text{\textit{kl}}}\left(\pmb{s},\pmb{\text{\textit{$s'$}}}\right)=(2 \pi )^{-1} \sum _{p=0}^n \omega _{\text{\textit{kl}},p} \cos (p \chi )
\cos (p \pi /2)\int \kappa ^2\exp \left(-\alpha ^2\kappa ^2\right)J_p(\kappa  s)d\kappa \text{                    }(3.10)\text{        }\]

with $\alpha \approx $\(2\pi \left/s_0\right.\), \(s_0\) is the substrate lattice constant. Performing the $\kappa $ integral gives for the interaction
between adatom monomers (\textit{ k,l}=1) or pairs (\textit{ k,l}=2,3) located at the origin and \(\pmb{s}\), using polar coordinates (\textit{ s},$\chi
$) for their distance \textit{ s} = $|$\(\pmb{s}\)$|$ and pair direction angle $\chi $ with respect to the (100) crystal axis is given by 

\[V_{\text{\textit{kl}}}(s,\chi )=\frac{1}{2\pi }\sum _p \omega _{\text{\textit{kl}},p} \frac{\left.\cos (\text{p$\chi $})\cos (\frac{p\pi }{2}\right)\Gamma
\left(\frac{p+3}{2}\right)s^p\, _1F_1\left(\frac{p+3}{2};p+1;\frac{-s^2}{4 \alpha ^2}\right)}{2^{p+1}\Gamma (p + 1)\alpha ^{p+3}}\text{  },\text{
   }(3.11)\]

where \(\, _1F_1\) denotes the Hypergeometric Function, \(\Gamma (p)\) the Gamma function, and the \(\omega _{\text{\textit{kl}},p}\) denote coefficients
of the cosine series Eq. (3.7). 

The shape of the interaction elements \(U_{\text{\textit{kl}}}\) depends strongly on sign and magnitude of the \(\omega _{\text{\textit{kl}},p}\).
\(U_{kl}(s,\chi )\) is finite at small distances \textit{ s} and approaches zero with { }

\[V_{\text{\textit{kl}}}(s,\chi ) =(2\pi )^{-1}\sum _p \omega _{\text{\textit{kl}},p} \cos (p \chi )\cos (p \pi /2)(p+1)(p-1)s^{-3}\text{  },\text{
   }s>>s_0 .\text{          }(3.12)\]

The eigenvalues \(\omega _{\text{\textit{kl}}\text{\textit{$,$}}p}\) as defined by the linear equations (3.4) and (3.5) are proportional to the product
of stress parameters \(P_k\)\(P_l\) and inverse proportional to the elastic constant \(c_{44}\) 

\[\omega _{\text{\textit{kl}},p}\text{  }\sim \text{  }P_kP_l/c_{44}\text{  }.\text{                                                            
                    }(3.13)\]

\subsection*{3.3.Elastic isotropic substrate}

For elastic isotropic substrates (\(c_{11}\)-\(c_{12}\)-2\(c_{44}\)=0, Voigts notation) some eigenvalues { }\(\omega _{\text{\textit{kl}}\text{\textit{$,$}}p}\)
can be calculated { }in closed form [9]. They are listed in Tab.1 together with those calculated numerically for Tungsten as substrate . In the case
of pair interaction\textit{  k=l}=1 the force dipole tensors are isotropic and the eigenvalues for \textit{ p}$>$0 vanish. In the case of quattro
interaction \textit{ k=l}=2 the force dipole tensor is anisotropic and directed in the x-direction and the 3 eigenvalues differ. In the case of \textit{
k=l}=3 the force dipole tensor is directed in the y-direction and { }\(\omega _{\text{\textit{kl}},2}\) changes sign. For the case of trio interactions
\textit{ k=1, l}=2 and \textit{ k=1, l}=3 no closed solution has been found. The closed solutions for elastic isotropic substrates are useful for
testing the code in the more general anisotropic case. { }

\[\begin{array}{|c|c|c|c|c|}
\hline
 \text{\textit{$k$}} & \text{\textit{$l$}} & \omega _{\text{\textit{kl}},0} & \omega _{\text{\textit{kl}},2} & \omega _{\text{\textit{kl}},4} \\
\hline
 1 & 1 & -A & 0 & 0 \\
\hline
 1 & 2 & -0.509 & -0.509 & 0. \\
\hline
 1 & 3 & -0.509 & 0.509 & 0. \\
\hline
 2 & 2 & -B & -A & C \\
\hline
 2 & 3 & C & 0 & -C \\
\hline
 3 & 3 & -B & A & C \\
\hline
\end{array}\]

Table 1. Coefficients \(\omega _{\text{\textit{kl}},p}\) (in \(P_k\)\(P_l\)/\(c_{44}\) units) { }for W(100), legend\\
\hspace*{0.5ex} A=\(\text{  }c_{11}\)/(2(\(c_{11}\)-\(c_{44}\)))\\
\hspace*{0.5ex} B= { }(\(5c_{11}\)-2\(c_{44}\))/(8(\(c_{11}\)-\(c_{44}\)))\\
\hspace*{0.5ex} C= { }(\(c_{11}\)-2\(c_{44}\))/(8(\(c_{11}\)-\(c_{44}\))).\\

The elastic constants used were \(c_{11}\)=523, \(c_{12}\)=203, \(c_{44}\)=160 GPa.\\
From Tab.1 and Eq. (3.12) we conclude the following long range interactions if \(P_1\)$>$0 and \(P_2\)$>$0:\\
- the pair interaction \(V_{11}\) is repulsive\\
- the trio interaction \(V_{12}\) is repulsive in x-, weakly repulsive in 45${}^{\circ}$ direction and attractive in y-direction with half strength\\
- the quattro interaction \(V_{22}\) is strongly repulsive in x-, weakly attractive in 45${}^{\circ}$ direction and weakly attractive in y-direction\\
- the quattro interaction \(V_{23}\) is attractive in x- and y-direction and repulsive in 45${}^{\circ}$ direction.\\
Short range interactions from Eq. (3.11) differ from the simple long range scheme. 

\subsection*{3.4.Elastic anisotropic substrate}

For elastic anisotropic substrates Eqs. (3.4) and (3.5) have to be solved numerically for all combinations $\{$\textit{ k,l}$\}$. The \(q_m(\pmb{\kappa
})\) are derived in Eq. (3.4) from the characteristic polynomial of 6th degree, keeping the 3 solutions with Re(\(q_m\))$<$0. The eigenvalues \(\omega
_{\text{\textit{kl}}}^{-1}(\pmb{\kappa })\) are then derived from Eq. (3.5). For the Pd(100) example we refer to eigenvalues presented in [6].

\section*{4.Discussion}

The current paper presents an extension of an earlier theory [3]. While previously the adatoms were treated as monomers interacting via substrate
strain, the extension allows accounting for dimer effects on adatom configurations. 

The model is based on the theory of elasticity in the substrate and on the lateral stress adatoms apply to the surface. Key assumption is the mechanism
by which adatoms and dimers interact. Monomer adatoms sitting on fourfold (or threefold) adatom sites expand or contract the substrate by creating
isotropic stress. The well-established effect of substrate strain generated by lattice misfit between substrate and overgrown is approximated by
the anisotropic stress adatom pairs exert to the substrate. The latter mechanism should work in cases where adatom pairs are bound electronically
and stretched (or compressed) due to their position on substrate sites.

The current model neglects other interactions like dipole-dipole repulsion.

An elastic continuum model for the substrate is rated inadequate for describing short range effects, but describes well long range effects and elastic
anisotropies (strong in case of adatom dimers). The cutoff mechanism used in Eq. (3.9) avoids a { }\(s^{-3}\) singularity and therefore extends the
models reach towards smaller distances. A different cutoff mechanism will lead to interactions different at small distances. A hard cutoff e.g. will
lead to a damped oscillating interaction.

The restriction to high symmetry adatom locations has the advantage of stress parameter \(P_k\) degeneracy. On (100) surfaces \(P_3\)=\(P_2\) due
to equivalence of x-directed and y-directed dimers. On (111) surfaces \(P_4\)=\(P_3\)=\(P_2\) for similar symmetry reasons. This reduces the number
of free parameters of the model. The dependence of stress parameter values \(P_k\) from coverage could be an issue. Its value for isolated dimers
could differ from its value in islands since the bond between adatoms depends on their coordination. Different species of adatoms may show different
dependencies which could be analyzed with the help of a density-functional-theory calculation. { } 

A further key assumption is an ideal flat surface, i.e. the absence of steps which are known for their significant attractive or repulsive interaction
with adatoms. 

Unknown is the magnitude of the stress parameters \(P_k\). This is the reason for presenting the interactions \(V_{\text{\textit{kl}}}\) in scaled
units only. It is, however, assumed that the qualitative statements on the directional dependencies and on the interactions sign hold.

The tungsten example as substrate material was chosen because tungsten is elastically isotropic and therefore allows to test the code for numerical
calculations of the eigenvalues $\omega $. The question, if elastic interactions are relevant for interpreting first principles calculations, is
left to a forthcoming paper [6].

\section*{7.Summary}

A model for substrate strain mediated monomer adatom interaction is extended to comprise dimers. While monomers sitting on high symmetry sites exert
isotropic stress to the substrate the stress of dimers is strongly anisotropic which causes qualitatively different strain fields and interactions.
The interaction of dimers is denoted quattro interaction, the interaction of a dimer with a monomer is denoted trio interaction. The model is evaluated
by establishing and numerically solving elastic plane wave eigenvalue equations. For the elastically isotropic tungsten substrate a closed solution
for quattro interactions is given, for trio interactions numerical values are presented.

\section*{References}

[1] T.L.Einstein, Interactions between Adsorbate Particles, in: Physical Structure of Solid Surfaces (W.N.Unertl ed.), Elsevier, Amsterdam (1996)\\

[2] K.H.Lau, W. Kohn, 1977, Surface Sci. 65, 607 (1977)\\
 { }

[3] W.Kappus, Z.Physik B 29, 239 (1978) \\

[4] K.H.Lau, Sol. State Commun. 28, 757 (1978)\\

[5] Y.Zhang, V.Blum, K.Reuter, Phys. Rev. B 75, 235406 (2007)\\

[6] W.Kappus, to be published\\

[7] P.M{\" u}ller, A.Saul, Surf. Sci. Rep. 54, 157 (2004)\\

[8] H.Horner, H.Wagner, Adv. in Phys. 23,587 (1974)\\

[9] W.Kappus, Z.Phyisik B-Condensed Matter 38, 263 (1980)\\

\section*{Acknowledgement}

Feedback to this manuscript would be gratefully acknowledged.

\section*{Figures}

\includegraphics{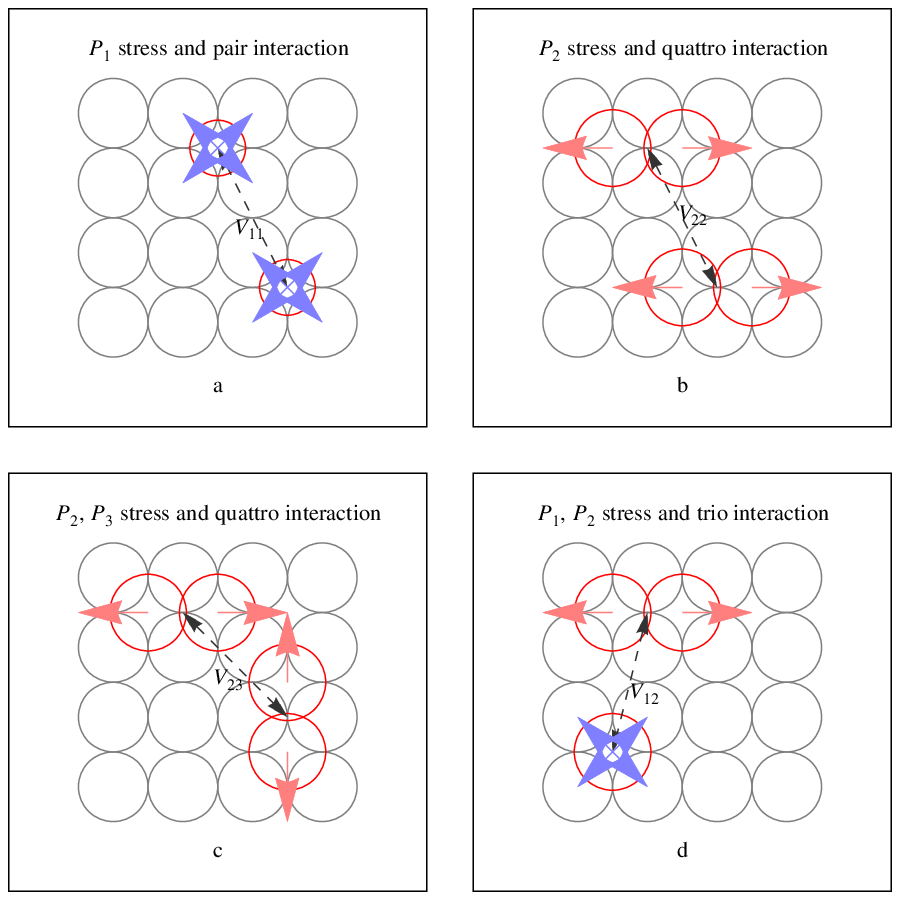}

Fig.1: (red) adatoms on (gray) substrate positions with associated stress tensors and interaction types on (100) surfaces. Stress tensors \(\pmb{P}_1\)
are depicted by blue arrows. Stress tensors\(\pmb{ }\pmb{P}_2\) and \(\pmb{P}_3\) are depicted by red arrows. { }Black dashed arrows indicate the
different interaction types \(V_{\text{\textit{kl}}}\) as defined in section 2.\\
a. monomers with stress tensor\(\pmb{ }\pmb{P}_1\) and interaction via \(V_{11}\). b. pairs with stress tensor \(\pmb{P}_2\) { }interacting via \(V_{22}\).
c. pairs with stress tensors \(\pmb{P}_2\) and \(\pmb{P}_3\) { }interacting via \(V_{23}\). d. monomer with stress tensor \(\pmb{P}_1\) { }and pair
with stress tensor \(\pmb{P}_2\) interacting via \(V_{12}\).

$\copyright $ Wolfgang Kappus (2016)

\end{document}